\documentstyle[12pt,epsfig,philmag,amstext]{article}

\newcommand{\gsurf}{\gamma_{\text{surf}}}
\newcommand{\gsf}{\gamma_{\text{sf}}}
\newcommand{\gus}{\gamma_{\text{us}}}

\begin{document}

\begin{center}
  {\Large\bf The influence of surface stress on dislocation emission
  from sharp and blunt cracks in f.c.c.\ metals}\\[\baselineskip]
  {\sc J. Schi{\o}tz}\\[0.5\baselineskip]
  {\small Center for Atomic-scale Materials Physics (CAMP) and\\
    Department of Physics, Building 307, Technical University of Denmark,\\
    DK-2800 Lyngby, Denmark}\\[0.5\baselineskip]
  and {\sc A. E. Carlsson}\\[0.5\baselineskip]
  {\small Department of Physics, Washington University, St.~Louis, MO
    63130, USA}\\[\baselineskip]
  (To be published in {\em Philos.~Mag.~A\/})\\[\baselineskip]
\end{center}

\begin{quote}
  \begin{center}
    {\sc Abstract}
  \end{center}
  \small We use computer simulations to study the behavior of
  atomically sharp and blunted cracks in various f.c.c.\ metals.  The
  simulations use effective medium potentials which contain many-body
  interactions.  We find that when using potentials repre\-sen\-ting
  platinum and gold a sharp crack is stable with respect to the
  emission of a dislocation from the crack tip, whereas for all other
  metals studied the sharp crack is unstable.  This result cannot be
  explained by existing criteria for the intrinsic ductile/brittle
  behavior of crack tips, but is probably caused by surface stresses.
  When the crack is no longer atomically sharp dislocation emission
  becomes easier in all the studied metals.  The effect is relatively
  strong; the critical stress intensity factor for emission to occur
  is reduced by up to 20\%.  This behavior appears to be caused by the
  surface stress near the crack tip.  The surface stress is a
  consequence of the many-body nature of the interatomic interactions.
  The enhanced dislocation emission can cause an order-of-magnitude
  increase in the fracture toughness of certain materials, in which a
  sharp crack would propagate by cleavage.  Collisions with already
  existing dislocations will blunt the crack, if this prevents further
  propagation of the crack the toughness of the material is
  dramatically increased.
\end{quote}

\begin{center}
  {\sc \S 1. Introduction}
\end{center}

One of the major problems in modeling fracture in materials is the
enormous range of length-scales that enter the problem (Carlsson and
Thomson 1998)\nocite{CaTh98}.  The atomic processes near the crack tip
require atomic-scale modeling, but the long-range elastic field of
a crack tip may interact with dislocations and dislocation sources too
far from the crack tip to allow treatment with atomic-scale methods,
except for a few highly idealized configurations.  More direct
interactions between crack tips and dislocations have also been
suggested, including the proposal that collisions between the crack
tip and dislocations may both be likely and be important for the
fracture toughness of real materials (Mesarovic 1997)\nocite{Me97}. 
Here we concentrate on the atomic-scale behavior of the crack
tip, especially in the context of this proposal.

The main conclusion of Mesarovic's work is that the elastic field of a
moving crack will attract preexisting dislocations towards the crack,
causing them to collide with the crack or at least to pass so close to
the crack that an oppositely signed dislocations will be emitted.  This
will usually cause local blunting of the crack front.  Mesarovic
assumes that blunting a segment of the crack front causes that segment
to arrest, forcing the remainder of the crack to move around the
arrested crack.  This causes the {\em local\/} crack tip toughness
$(\Gamma_{\mbox{tip}})$ to be increased significantly.  A positive
feedback mechanism then ensues. The increased crack tip toughness
causes a larger stress in the surrounding material, increasing the
activity of the dislocations and dislocation sources, thus further
raising the fracture toughness by a dramatic increase in the shielding
of the crack (Beltz, Rice, Shih and Xia 1996; Mesarovic
1997)\nocite{BeRiShXi96,Me97}.

This model clearly depends on the ability of a single dislocation to
arrest the crack by blunting it.  It thus becomes important to
understand how crack blunting affects the intrinsic behavior of the
crack tip.  By analogy to the stress field near an elliptic crack, it
is generally assumed that blunting a crack will dramatically reduce
the stress concentration at the crack tip, and thus make both crack
propagation and dislocation emission much harder even after a single
layer of blunting (Paskin, Massoumzadeh, Shukla, Sieradzki and Dienes
1985; Gumbsch 1995)\nocite{PaMaShSiDi85,Gu95}.  In a previous paper
(Schi{\o}tz, Canel and Carlsson 1997)\nocite{ScCaCa97} we found that
not to be the case.  We studied the effect of blunting on subsequent
dislocation emission or cleavage for a crack in a two-dimensional
hexagonal lattice, with the interatomic interactions described by a
simple pair potential.  We reached two main conclusions: First, the
effects of the crack blunting on the elastic fields around the crack
are minimal, even at very short ranges.  This is because the
singularity of a 60$^\circ$ wedge crack is almost as strong as for a
sharp crack (the power is $-0.488$ compared to $-0.5$).  Secondly, the
non-linearities in the interatomic interactions cause the blunting to
have some effect: a slight increase in the stress intensity factor
required to cleave, and a tendency to favor emission in stead of
cleavage as soon as a single layer of blunting is introduced.

Since these effects are dependent on the interatomic potential used, we
decided to perform atomic-scale simulation of crack blunting using more
realistic many-body potentials for selected f.c.c.\ metals (Ni, Cu, Pd,
Ag, Pt, Au).  These simulations are described in this paper.  Rather
than attempting an exhaustive set of simulations of all likely crack
shapes, we have focussed on a single shape (the shape previously
studied), and focus our attention on the effects of including
many-body interactions in simulations of this particular shape.  As we
shall see, the effects we observe will apply for a large class of crack
shapes.

We find that the inclusion of many-body effects in the interatomic
potential dramatically enhances the effects of crack blunting.  This
is mainly caused by the introduction of surface stresses, which are
absent in models based on a nearest-neighbor pair potential.  For most
materials surface stresses are tensile\footnote{We here use the
  convention that the surface stress is \emph{tensile} if the surface
  is under tension, i.e.\ if the energy of the \emph{surface} atoms
  could be reduced by reducing the lattice constant.  Unfortunately
  the opposite definition is sometimes also seen.}, and a tensile
stress in the surface at the end of a blunt crack will increase the
resolved shear stress on a dislocation about to be emitted, thus
enhancing dislocation emission.  We see reductions of the stress
intensity factor required for dislocation emission of up to 20\%
caused by just a few layers of blunting, this is enough to change the
behavior of the crack from brittle crack propagation to dislocation
emission.

The simulations also provide an opportunity to test two recently
proposed criteria for the intrinsic behavior of sharp cracks (Rice
1992; Zhou, Carlsson, and Thomson
1994)\nocite{ZhCaTh94}\nocite{Ri92,ZhCaTh94}.  We find that neither
of the criteria are able to predict whether sharp cracks in the
investigated metals behave in an intrinsically ductile or brittle
manner.  The deviations seem to be caused by surface stresses in the
crack faces.  The ductility criterion proposed by Zhou \emph{et al.}
may be modified to take this into account.

\begin{center}
  {\sc \S 2. Computer simulations}
\end{center}

\begin{figure}[t]
  \begin{center}
    \leavevmode
    \epsfig{file=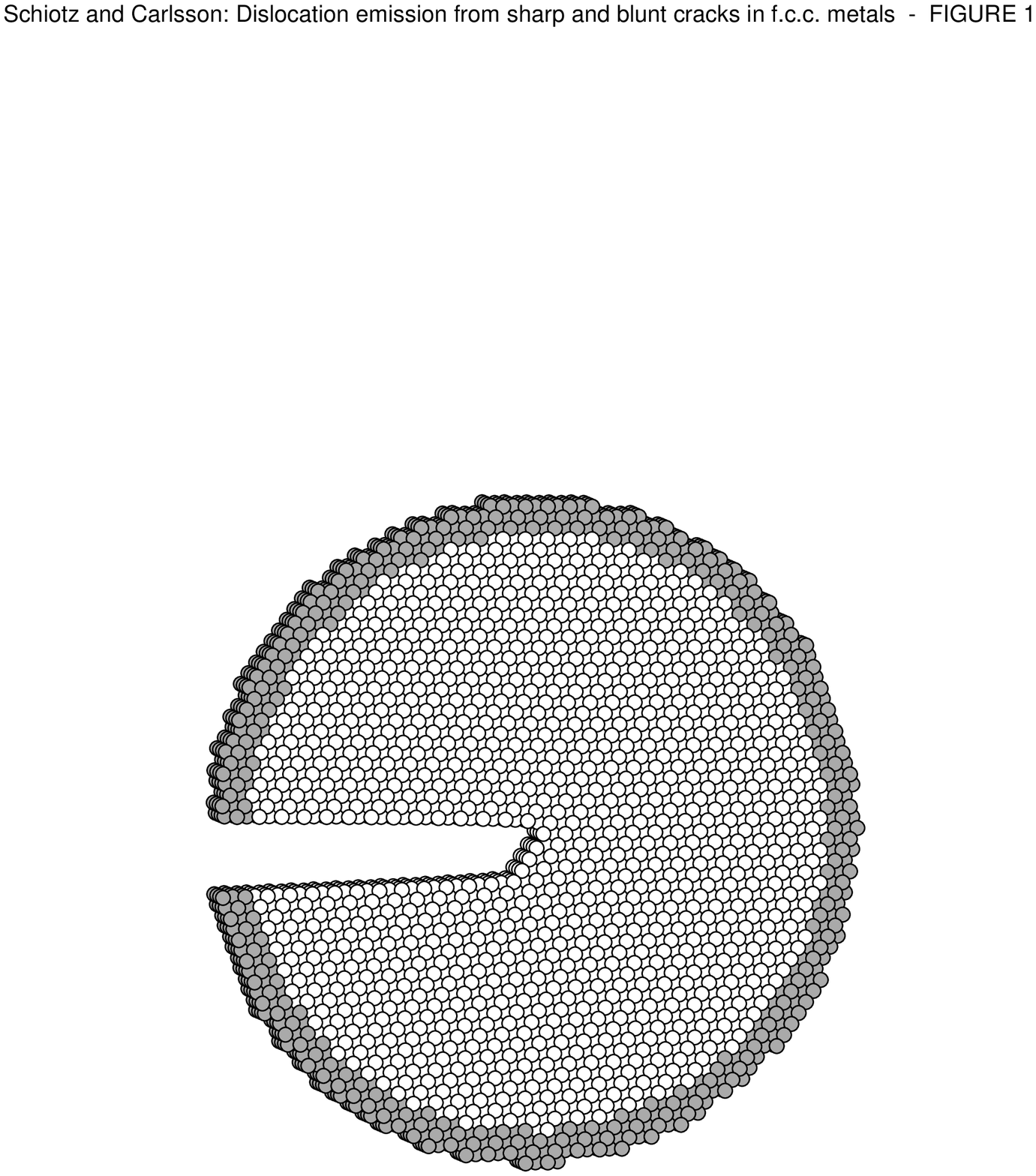, clip=, width=0.6\linewidth}
    \caption{\small The initial geometry of the simulation cell.  The grey
      atoms are static atoms. Their positions are fixed to match the
      displacement field of the crack as given by anisotropic linear
      elasticity (Sih and Liebowitz 1968)\protect\nocite{SiLi68}.  The white
      atoms are dynamic atoms, which are allowed to move. Their
      positions are energy-minimized for any given elastic load (given
      by the positions of the static atoms).  In the system shown here
      four layers of blunting are applied to the crack.}
    \label{fig:geo}
  \end{center}
\end{figure}
We have simulated the emission of the leading partial dislocation from
a crack using the geometry shown in figure \ref{fig:geo}.  The
simulation cell consists of a cylinder where the atoms near the
surface of the cylinder have fixed displacements (static atoms), and
the atoms inside the cylinder are free to move (dynamic atoms).
Periodic boundary conditions are applied in the direction along the
cylinder axis (the $z$-axis).  We create a sharp crack by first
displacing all the atoms according to the displacement field of a
crack loaded at the Griffith criterion.  We then minimize the energy
with respect to all coordinates of the dynamic atoms to investigate if
the sharp crack is stable.  To create a blunt crack we proceed in a
similar fashion.  First we remove one to five half-layers of atoms to
create the crack.  Then we displace all the atoms.  The displacement
field around a blunt crack is not known analytically, but at large
distances compared to the blunting the displacement field will be
identical to that of a sharp crack.  We therefore use that expression
for the field; any error introduced near the crack tip will quickly be
removed by the minimization algorithm.  We can then gradually step up
the loading of the crack, until the leading partial dislocation is
emitted or the crack cleaves.

\begin{figure}[t]
  \begin{center}
    \leavevmode
    \epsfig{file=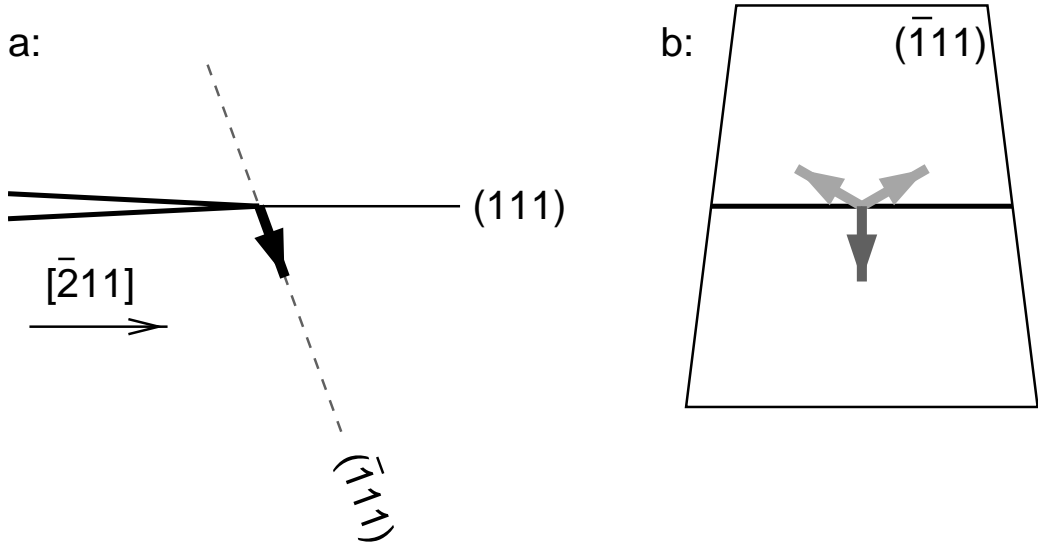, clip=, width=0.9\linewidth}
  \end{center}
  \caption{\small The orientation of the crack.  (a) Side view.  The crack is
    moving on the $(111)$ plane in the $[\bar{2}11]$ direction.  A
    partial dislocation can be emitted on the $(\bar{1}11)$ plane; the
    thick arrow shows its Burgers vector.  The $(111)$ plane is not a
    mirror plane. Emission in the upward direction would have to
    happen in the same $(\bar{1}11)$ plane, and thus in a
    ``backwards'' direction with respect to the crack.  (b) Front view
    of the crack showing the Burgers vectors of the possible partial
    dislocations.  The dislocations that can be emitted upward
    (lighter grey) have a large screw component in their Burgers
    vector, and therefore do not couple as strongly to the stress
    field of the mode I crack.}
  \label{fig:orientation}
\end{figure}
We have chosen a $(111)[2\bar{1}\bar{1}]$ orientation of the crack,
i.e.\ the crack moves along a $(111)$ plane in the $[\bar{2}11]$
direction.  Partial dislocations can be emitted on the $(\bar{1}11)$
plane.  Since the $(111)$ plane is not a mirror plane there is only
one slip plane, but dislocations can be emitted in both directions.
Linear elasticity favors emission in the ``forward'' direction with
respect to the crack.  The leading partial dislocation emitted in the
favored ``forward'' direction has pure edge character, whereas the
leading partial dislocation that could potentially be emitted in the
backwards direction has a large screw component, and therefore couples
less strongly to the stress field under mode I (pure opening) loading.
See Figure~\ref{fig:orientation}.  If the crack direction were
reversed (i.e.\ $[2\bar{1}\bar{1}]$ instead of $[\bar{2}11]$) the
situation would be reversed, and the system would have the choice
between emitting an unfavorable screw-like dislocation in the
favorable forward dislocation, or a favorable dislocation in an
unfavorable direction.  This would lead to a suppression of the
emission and a more brittle behavior.

\begin{table}[t]
\newcommand{\mbr}[1]{#1}
\newcommand{\expt}{\multicolumn{2}{r|}{Experiment:}}
\newcommand{\abinit}{\multicolumn{2}{r|}{{\em ab initio\/}:}}
\begin{center}
  \begin{tabular}[c]{lr|cccccc}
      \hline\hline
      \multicolumn{2}{c|}{Quantity} &
                                 Ni   &  Cu   &  Pd   &  Ag   &  Pt   &  Au \\
      \hline                 
      $d$ & \mbr{(\AA)}       & 3.491 & 3.592 & 3.878 & 4.063 & 3.921 & 4.055\\
      $C_{11}$ & \mbr{(GPa)}  & 242.9 & 172.8 & 217.6 & 125.7 & 319.0 & 196.7\\
      \expt                   & 261.2 & 176.2 & 234.1 & 131.5 &  ---  & 201.6\\
      $C_{12}$ & \mbr{(GPa)}  & 150.8 & 116.0 & 162.3 & 87.49 & 258.9 & 162.6\\
      \expt                   & 150.8 & 124.9 & 176.1 & 97.3  &  ---  & 169.7\\
      $C_{44}$ & \mbr{(GPa)}  & 147.5 & 90.60 & 75.42 & 54.57 & 80.23 & 46.77\\
      \expt                   & 131.7 & 81.8  & 71.2  & 51.1  &  ---  & 45.4\\
      \hline
      $\gsurf^{111}$ & \mbr{$(J/m^2)$}  &
                                1.684 & 1.036 & .6677 & .5472 & .7492 & .5143\\
      \abinit                 & 2.011 & 1.952 & 1.920 & 1.172 & 2.299 & 1.283\\
      $\gsurf^{100}$ & \mbr{$(J/m^2)$}  &
                                1.800 & 1.125 & .7663 & .6145 & .8766 & .5990\\
      \abinit                 & 2.426 & 2.166 & 2.326 & 1.200 & 2.734 & 1.627\\
      $f^{111}$&\mbr{$(J/m^2)$}&-0.30 & 0.265 & 1.104 & 0.486 & 1.941 & 1.291\\
      \abinit                 &  ---  &  ---  & 3.685 &  ---  & 6.761 & 2.772\\
      $f^{100}$&\mbr{$(J/m^2)$}&1.041 & 1.017 & 1.328 & 0.788 & 1.978 & 1.327\\
      \abinit                 &  ---  &  ---  & 2.211 & 1.682 & 5.656 & 3.140\\
      $\gsf$ &\mbr{$(mJ/m^2)$}& 54    & 31    & 6.4   & 8.2   & 7.4   & 8.0\\
      \abinit                 & 187   &    56 &   225 &    34 &   393 & 59\\
      $\gus$ &\mbr{$(mJ/m^2)$}& 278.2 & 173.6 & 147.6 & 114.2 & 160.9 & 100.4\\
      $K_G$ & \mbr{$(MJ m^{-5/2})$} &
                                .9825 & .6194 & .4843 & .3612 & .5479 & .3476\\
      \hline
      \multicolumn{2}{l|}{$\gus / \gsurf^{111}$} &
                                .165  & .168  & .221  & .209  & .215  & .195\\
      \multicolumn{2}{l|}{$\gus / \mu b$} &
                                .0182 & .0180 & .0166 & .0170 & .0167 & .0174\\
      \multicolumn{2}{l|}{$\frac{\gus}{\mu b} \left(1 + \frac{f^{111}}{3 \gsurf^{100}}\right)$} &
                                .0172 & .0194 & .0246 & .0215 & .0290 & .0299\\
      \hline\hline
    \end{tabular}
  \end{center}
  \caption{\small Elastic constants, surface- and planar fault energies
    for six f.c.c\ metals in the Effective Medium Theory. $d$ is the
    lattice constant, $C_{11}$, $C_{12}$ and $C_{44}$ are the elastic
    constants.  $\gsurf^{111}$ and $\gsurf^{100}$ are the surface
    energies of the (111) and (100) surfaces.  $f^{111}$ and $f^{100}$
    are the surface stresses.  $\gsf$ and $\gus$ are the stacking
    fault energy and the unstable stacking energy.  $K_G$ is the
    Griffith critical stress intensity factor in plane strain for the
    crack orientation under consideration, calculated from
    $\gsurf^{111}$ and the elastic constants.  Experimental values:
    elastic constants are from Kittel (1996)\protect\nocite{Ki96}.
    \emph{Ab initio} calculations: $\gsurf$ from Vitos, Ruban, Skriver
    and Koll\'ar\protect\nocite{ViRuSkKo98}; $f^{111}$ for Pt and Pd
    from Feibelman (1995)\protect\nocite{Fe95}, Au(111) from Needs and
    Mansfield (1989)\protect\nocite{NeMa89}; $f^{100}$ from
    Fiorentini, Methfessel and Scheffler
    (1993)\protect\nocite{FiMeSc93}; $\gsf$ from Rosengaard and
    Skriver (1993)\protect\nocite{RoSk93}.}
  \label{tab:emt}
\end{table}
The interatomic interactions are described by the Effective Medium Theory
(Jacobsen, N{\o}rskov and Puska 1987; Jacobsen, Stoltze and N{\o}rskov
1996)\nocite{JaNoPu87,JaStNo96}.  The main materials parameters for
the simulated materials are given in table~\ref{tab:emt}.  All values
calculated are given for the equilibrium lattice constants of the
EMT-metals.  The equilibrium lattice constants differ {\em slightly\/}
from the values used to construct the original EMT parameters, since
the extension of the interactions beyond nearest neighbors (Stoltze
1990; Jacobsen {\em et al.} 1996)\nocite{St90,JaStNo96} causes a
contraction of the lattice of around 0.5\%.  It is seen that the
elastic constants are close to the experimental values, but that the
surface energies and stresses are quite low.

The surface stresses presented in table~\ref{tab:emt} were calculated
from the difference in the energy variation versus strain of a slab
and a bulk system.  Care was taken to ensure convergence with respect
to system size.  As a test, the surface stresses were also calculated
using an atomic-level definition of the stress tensor (Egami, Maeda
and Vitek 1980; Ray and Rahman 1984)\nocite{EgMaVi80,RaRa84}.  The two
calculations agree.

\begin{center}
  {\sc \S 3. Results}
\end{center}

\begin{figure}[t]
  \begin{center}
    \leavevmode
    \epsfig{file=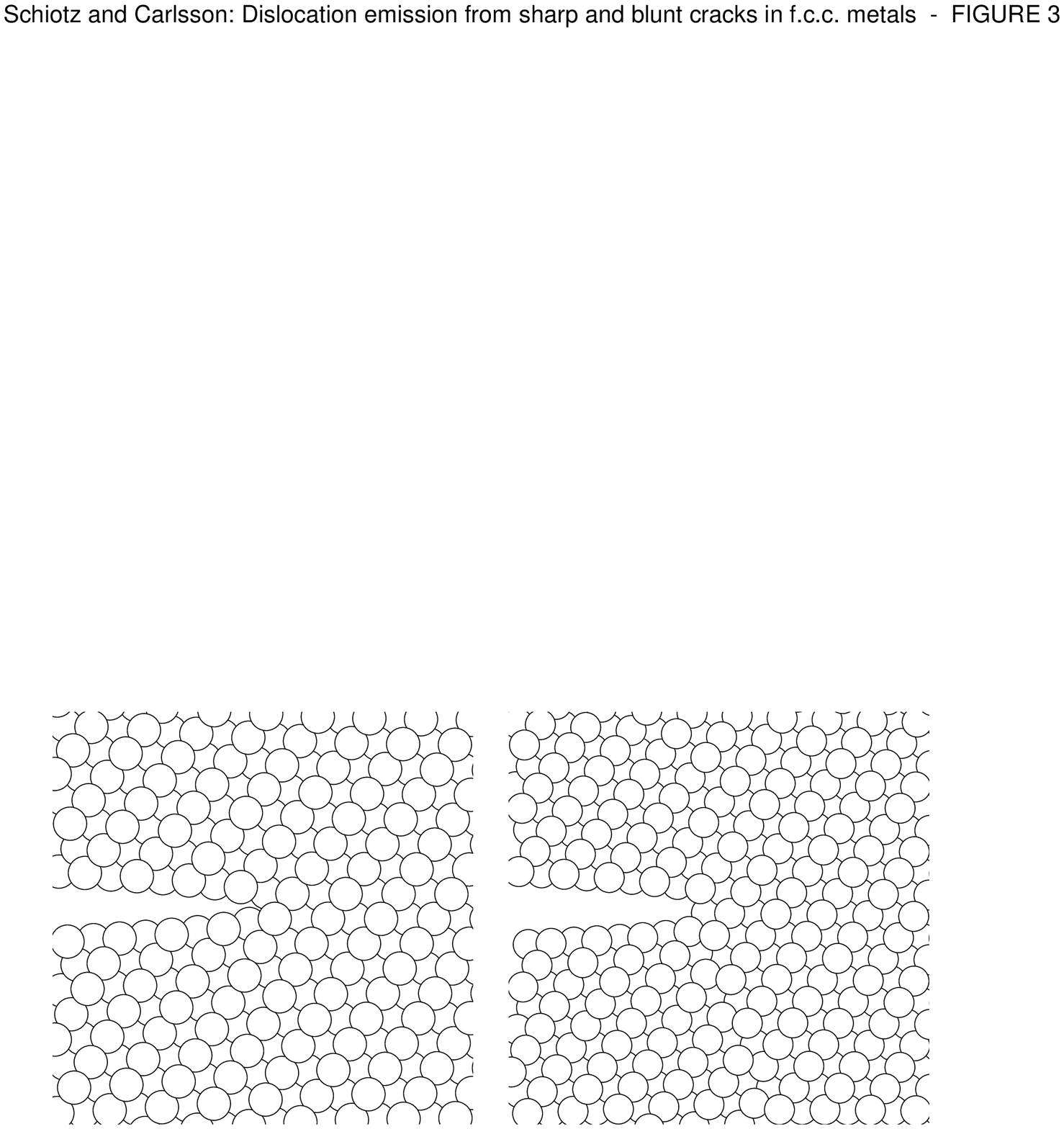, clip=, width=0.95\linewidth}
    \caption{\small The behavior of sharp cracks.  {\em Left:\/} A sharp crack
      in gold is stable. No dislocation is emitted.  {\em Right:\/} A
      sharp crack in nickel is unstable. A partial dislocation is
      emitted (down and to the right).}
    \label{fig:sharp}
  \end{center}
\end{figure}

\begin{table}[t]
  \begin{center}
    \begin{tabular}[c]{c|cccccc}
      \hline\hline
      Blunting &  Ni  &  Cu  &  Pd  &  Ag  &  Pt  &  Au \\
      \hline
      & \multicolumn{6}{c}{\underline{Cleavage of sharp crack:}}\\
      0        & {\em unstable} & {\em unstable} & {\em unstable} &
      {\em unstable} & 0.98--1.07 & 0.98--1.05 \\
      & \multicolumn{6}{c}{\underline{Emission from blunt crack:}}\\
      1        & 0.85 & 0.87 & 1.01 & 0.94 & 1.05 & 1.03 \\
      2        & 0.80 & 0.80 & 0.86 & 0.83 & 0.86 & 0.83 \\
      3        & 0.80 & 0.78 & 0.82 & 0.81 & 0.82 & 0.79 \\
      4        & 0.81 & 0.79 & 0.82 & 0.82 & 0.81 & 0.77 \\
      5        & 0.82 & 0.80 & 0.82 & 0.83 & 0.81 & 0.78 \\
      \hline\hline
    \end{tabular}
  \end{center}
  \caption{\small Critical load as a function of crack blunting, given in
    units of the Griffith load ($K_G$).  First line (blunting = 0):
    The interval where the sharp crack is stable.  For loads above the
    given interval the crack advances, below it it retreats.  For some
    metals a sharp crack is unstable and spontaneously emits a
    dislocation.  The following lines (blunting $>$ 0): The load at
    which the leading partial dislocation is emitted.}
  \label{tab:results}
\end{table}
The results of the simulations are shown in table \ref{tab:results}.
There are two main results.  First, the simulations of the sharp crack
show that the investigated FCC metals fall in two groups: platinum and
gold are intrinsically brittle (a sharp crack is stable, and will
propagate), whereas nickel, copper, palladium and silver are
intrinsically ductile (a sharp crack is unstable, and will emit a
dislocation).  See Figure \ref{fig:sharp}.  We find a small amount of
lattice trapping for platinum and gold, approximately 5\%.  We see
from the simulations that palladium and silver are very close to the
transition between intrinsically brittle and ductile behavior.  For
these two metals a sharp crack loaded at the Griffith load will remain
stable, but when the load is increased the crack will emit a
dislocation before the lattice trapping is overcome.

The second main result is the behavior of the blunt crack.  Here all
six metals behave in a similar way.  After just one layer of blunting
the crack will no longer cleave, but emits a dislocation.  Further
blunting of the crack seems to favor the emission further, at least
until the crack is blunted by three layers.  Increasing the blunting
beyond three layers has no or little effect on the emission.  For the
largest amount of blunting investigated (four and five layers of
blunting) we find a slight reversal of the effect - increasing the
blunting seems to increase the load required to emit a dislocation a
little.  The total reduction of the load required to emit a
dislocation is around 20\% compared to the Griffith load.

\begin{center}
  {\sc \S 4. Discussion}
\end{center}

The results presented in the previous section show both major
similarities to and differences from our previously published results
for blunt cracks in a two-dimensional lattice (Schi{\o}tz {\em et
  al.\/}~1997).  The main similarity is the change in behavior of the
crack for some force law parameters, a change from brittle behavior of
the sharp crack to ductile behavior of the blunt crack.  The main
difference is the ease with which the dislocations are emitted from the
blunt crack.  In the previous work we saw an {\em increase\/} in the
critical load when the blunting was increased, even when the crack began
emitting dislocations.  The emission was thus more a result of
inhibiting further cleavage than of enhancing the emission.  In this
work we instead see a clear {\em decrease\/} of the critical load.  We
further see that this decrease continues until the crack has been
blunted by three atomic layers.  This behavior is opposite from what one
would expect from simple arguments based on linear elasticity, since
increased blunting should decrease the stresses near the crack tip, and
thus increase the critical loads.  We should, however, remember that for a
blunt crack with sharp corners this {\em linear elastic\/} effect is
very small (Schi{\o}tz {\em et al.\/}~1997)\nocite{ScCaCa97}.

We find that the most probable cause of this difference is the
presence of surface stresses in these simulations.  The previous work
was done using nearest-neighbor pair potentials to describe the
interactions between the atoms; in such a potential no surface stress
appears.
The effective medium theory contains many-body interactions in the
potential, and gives a more realistic description of the metal
properties.  Surface stresses are present in the effective medium
theory.  As the crack is blunted, a new surface appears at the crack
tip, and a surface stress will be present at the new surface.

\begin{figure}[t]
  \begin{center}
    \leavevmode
    \epsfig{file=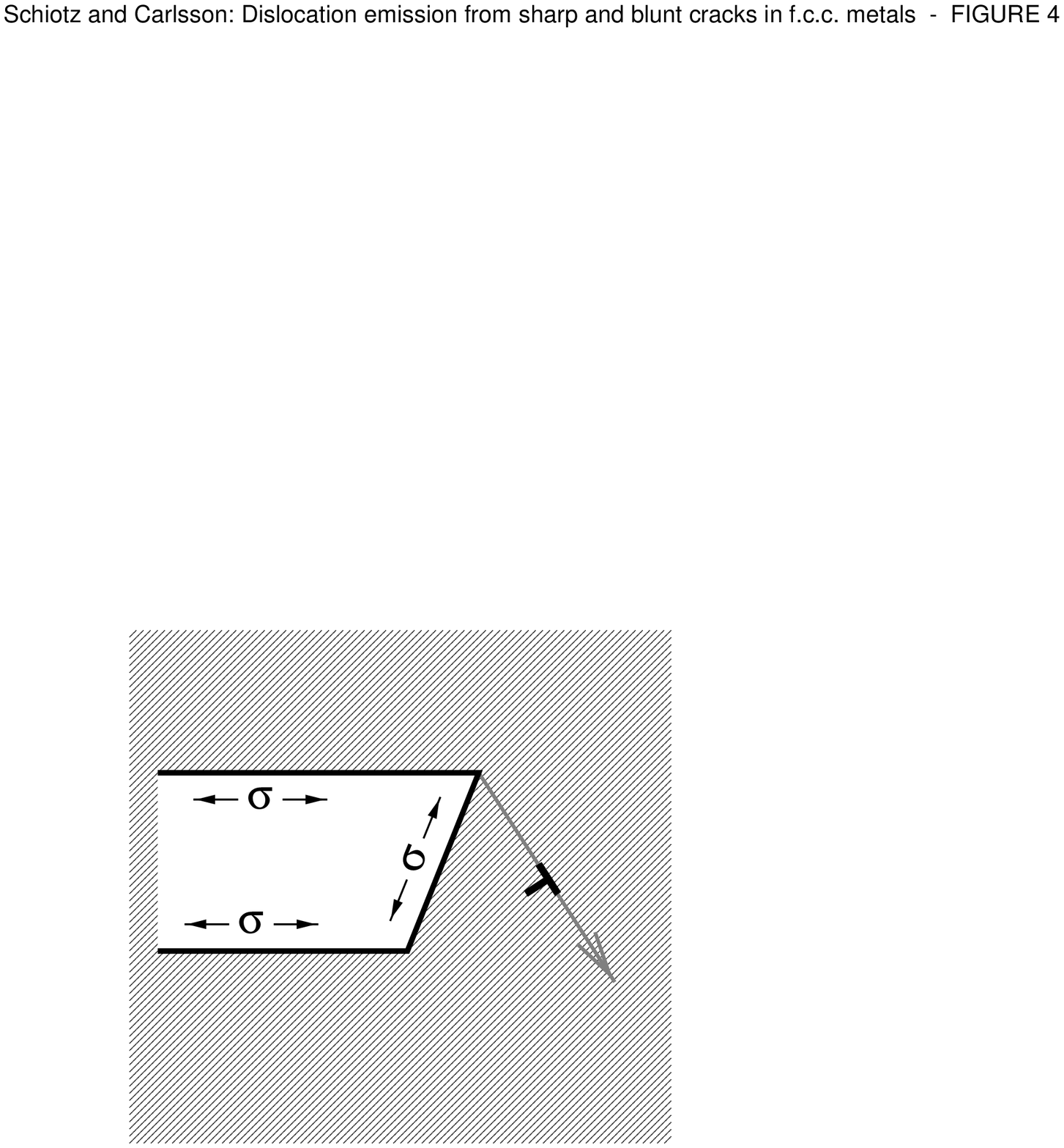, clip=, width=0.6\linewidth}
    \caption{\small Surface stresses near the tip of a blunt crack.  Tensile
      surface stress in the end surface of the crack will tend to
      favor dislocation emission on the slip plane shown.}
    \label{fig:crackstress}
  \end{center}
\end{figure}
\begin{figure}[t]
  \begin{center}
    \leavevmode
    \epsfig{file=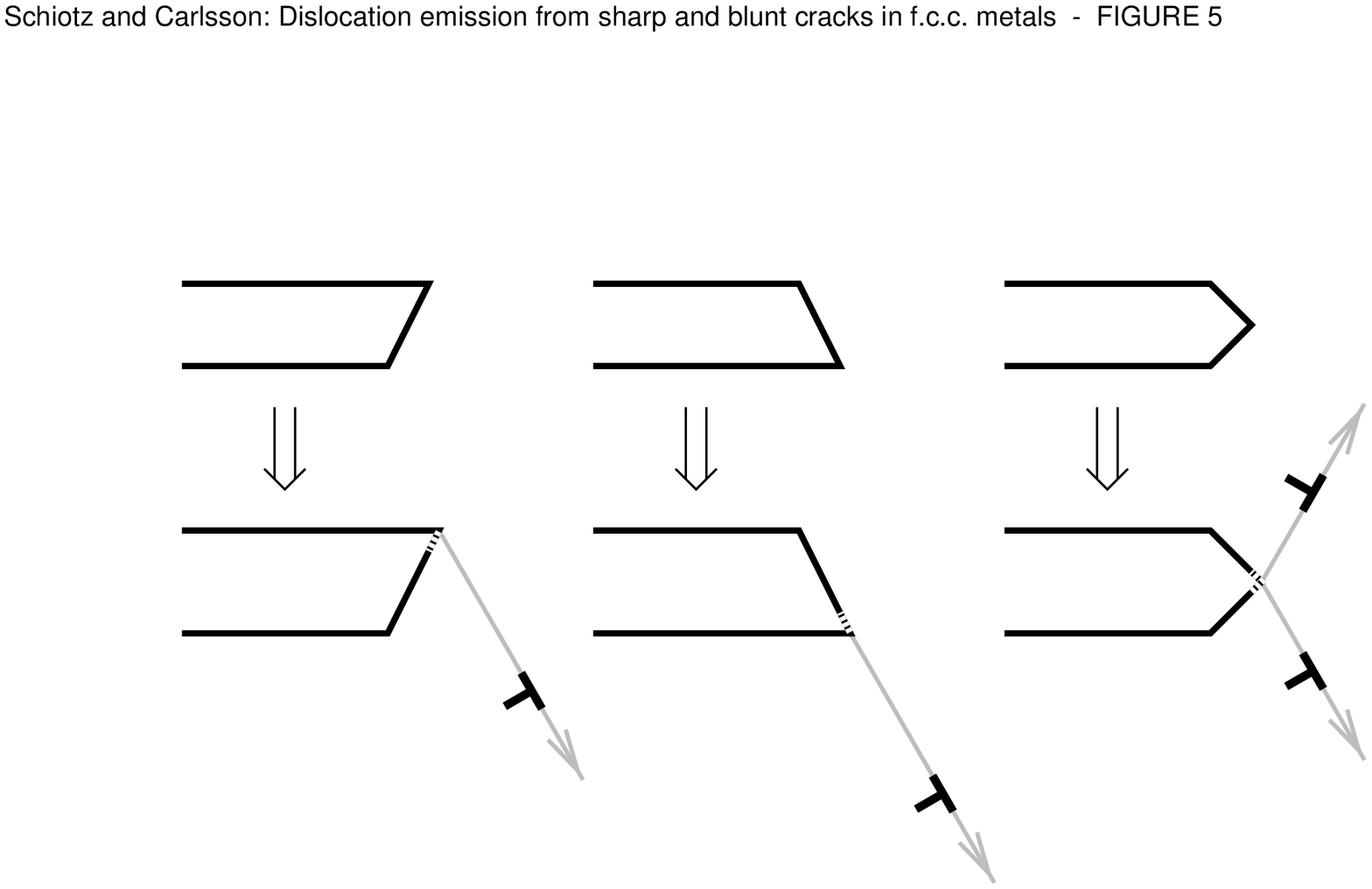, clip=, angle=-90, width=0.8\linewidth}
  \end{center}
  \caption{\small Dislocation emission from various hypothetical crack
    geometries.  In all cases the area of the ``end surface(s)'' of
    the crack is increased as a blunting dislocation is emitted.  A
    tensile surface stress will thus be relieved by the dislocation
    emission: the surface stress will facilitate dislocation emission.}
  \label{fig:othergeometries}
\end{figure}

For all the materials investigated here the surface stress of the
created $(100)$ facet is tensile.  This stress will naturally change
the stress distribution in the system near the crack tip.  Most
importantly, the tensile stress in the surface at the end of the crack
(see figure \ref{fig:crackstress}) will cause a resolved shear stress
in the plane of an emitted dislocation that favors emission.  This is
best seen by considering what emission of a dislocation will do to the
end surface under tensile stress.  As the dislocation is emitted the
end surface is enlarged by one row of atoms.  This allows the atoms to
contract, releasing some of
the tensile surface stress.  This energy argument is not limited to
the geometry studied here.  In most geometries of a blunt crack, there
will be an ``end surface'' which will become larger when a blunting
dislocation is emitted (see figure \ref{fig:othergeometries}).  A
tensile surface stress will thus be partially relieved by the emission
of a dislocation, and the surface stress must therefore aid the
emission.

As the size of the end ledge increases, the tensile stress builds
up\footnote{The surface atoms are lacking some of their neighbors
  compared to bulk atoms, the local electron density is thus lower.
  To compensate for this, it is energetically favorable to reduce the
  lattice spacing locally.  This is a slightly simplified picture of
  the main cause of the tensile surface stress.  When the ``new''
  surface at the end of the crack is small, i.e. when the blunting is
  low, the atoms will not lack all their neighbors, and the surface
  stress is expected to be lower.} and dislocation emission becomes
easier.  Once the end surface is sufficiently developed, the surface
stress reaches its ``macroscopic'' value, and further increase of the
crack blunting does not increase the surface stress.  No further
enhancement of the dislocation emission is then seen.  For this
reason, the critical load is approximately constant when the crack
blunting is changed from three to five layers (see
table~\ref{tab:results}).  It should be noted, that if the surface
stress is constant, the work done by it will to first order not depend
on the size of the end surface, nor by how the strain is distributed
along it.

To test this hypothesis we compare the shear stress induced by the
surface stress to the calculated reduction in the critical load
required to emit a dislocation.  An order of magnitude estimate of the
shear stress induced by the surface stress is $f^{100} / d$, i.e.\ the
surface stress is distributed over a layer of thickness comparable to
the lattice spacing.  The influence of this surface stress will then
depend on the ratio between this induced shear stress and the shear
stress originating from the loading of the crack.  The linear elastic
expression for the shear stress near a sharp crack loaded in mode I is
(Thomson 1986)\nocite{Th86}:
\begin{equation}
  \label{eq:modeIstress}
  \sigma_{r\theta} = K_I (2 \pi r)^{-1/2} \sin (\theta / 2) \cos^2
  (\theta / 2)
\end{equation}
This describes the stress field around a sharp crack. The $\theta$
dependence will change somewhat near a blunt crack, but as already
mentioned the $r$ dependence will be essentially unchanged.  The
linear elastic expression is only valid down to $r$ comparable to the
interatomic spacing, and at such distances it is only a rough
approximation.  Since we do not worry about factors of order unity, we
will nevertheless use Eq.~(\ref{eq:modeIstress}) at $r = d$ (the
lattice constant) as a rough estimate of the order of magnitude of the
shear stress induced near the crack tip by the loading of the crack.
For all the materials considered in this paper, the two estimates are
of the same order of magnitude (the ratio varying between 0.4 and
1.3).  It is thus plausible that the surface stress may have a
significant effect on the dislocation emission.

\begin{figure}[t]
  \begin{center}
    \leavevmode
    \epsfig{file=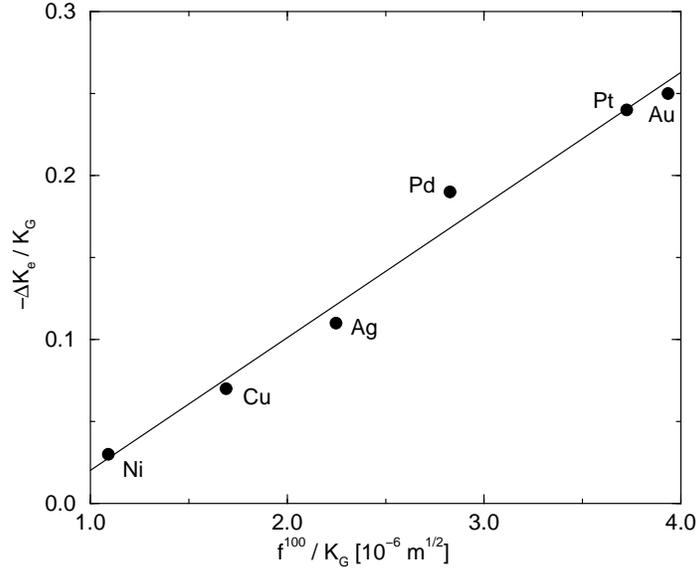, clip=, width=0.6\linewidth}
  \end{center}
  \caption{\small Comparison of the strength of the surface stress and the
    effect of blunting on the dislocation emission.  Along the
    $x$-axis is plotted the ratio between the surface stress
    ($f^{100}$) and the applied stress measured by the stress
    intensity factor ($K$).  Along the $y$-axis is plotted
    the enhancement of the dislocation emission as given by the change
    ($\Delta K_e$) in the load required to emit a dislocation measured
    in units of the Griffith stress.}
  \label{fig:correl}
\end{figure}
If the enhanced dislocation emission is indeed a result of the surface
stress, its magnitude should be roughly proportional to the surface
stress.  As a measure of the effect of the blunting we use the
relative change of the critical load to emit a dislocation as we
change the blunting from one to five layers ($\Delta K_e / K_G$).  In
Figure \ref{fig:correl} we investigate the correlation between the
reduction of the critical load ($-\Delta K_e / K_G$), and the ratio
between the surface stress ($f^{100}$) and the applied stress as
measured by the critical stress intensity factor ($K_G$).  The
remarkably good correlation is a strong support for the hypothesis
that the surface stress causes the enhanced dislocation emission.

Since a tensile surface stress will be present in most metals, 
this mechanism is possible in any metal where the sharp
crack behaves in an intrinsically brittle manner, but where the system
is not too far from dislocation emission.  In these materials
collisions between a propagating crack and preexisting dislocations
may cause the crack to locally become emitting, and if the collisions
are common enough the macroscopic response of the material may be
ductile in spite of the sharp crack tip being stable in principle.

\begin{figure}[tp]
  \begin{center}
    \leavevmode
    \epsfig{file=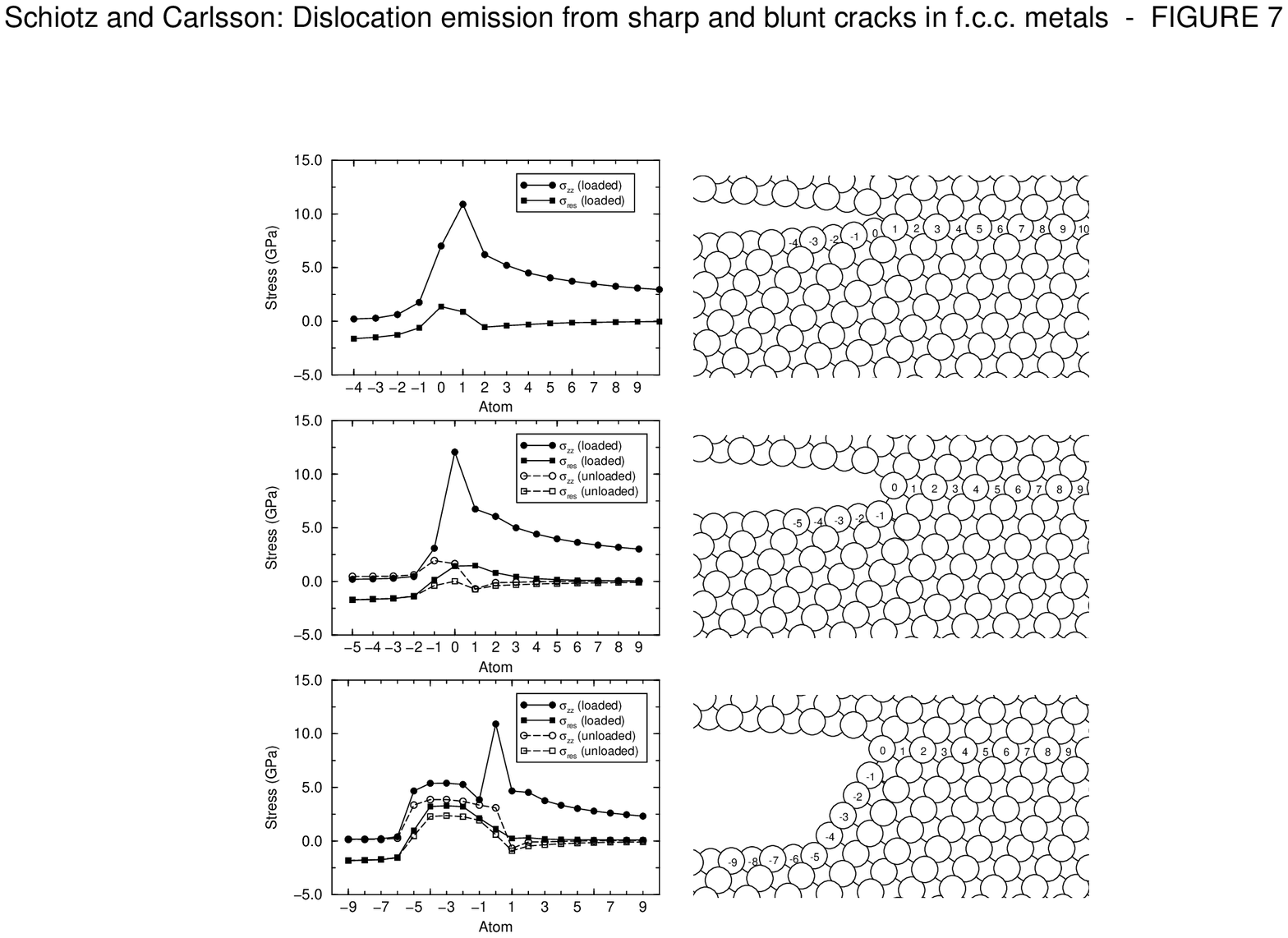, clip=, width=\linewidth}
  \end{center}
  \caption{\small Local stresses at the atoms near the crack tip in gold.
    The sharp crack is loaded at the Griffith load, the blunt cracks
    are loaded just below the load required to emit a dislocation.
    Filled circles show $\sigma_{zz}$, i.e.\ the component of the
    stress that drives the crack forward.  Filled squares show
    $\sigma_{\text{res}}$ (the resolved shear stress on the
    dislocation glide plane), i.e.\ the component of the stress
    driving dislocation emission.  Open symbols show the same
    quantities, but with no external load on the crack, so that the
    stresses shown are caused by the surfaces.  The sharp crack closes
    when not loaded, so no open symbols are shown in that case.}
  \label{fig:atomstress}
\end{figure}
Figure \ref{fig:atomstress} shows the stresses at selected atoms in
three different geometries in gold without loading, and loaded just
below the level required to emit a dislocation.  The open symbols are
the stresses without a load.  They give a measure of how much the
surface stresses contribute to the stresses driving the crack
($\sigma_{zz}$) and to the resolved shear stress on the dislocation
glide plane ($\sigma_{\text{res}}$).  It is seen that in the case of
five layers of blunting, the surface stresses contribute significantly
to both $\sigma_{zz}$ and $\sigma_{\text{res}}$ of the atoms near the
crack tip, but that the relative contribution to
$\sigma_{\text{res}}$ is larger.  This causes the surface stress to
enhance the dislocation emission more than the crack propagation.  It
should also be noted that {\em local} stresses near the crack tip may help
overcome barriers to crack propagation (lattice trapping), but can not
help propagating the crack, as crack propagation is energetically
forbidden below the Griffith loading.

Atomic-scale stresses such as those plotted in
Figure~\ref{fig:atomstress} should be viewed with some scepticism, as
there is no unique definition of the atomic-level stress.  Although
the different definitions give the same answer when averaged over a
small region of space, significant differences can be seen in the
values for individual atoms (Cheung and Yip 1991)\nocite{ChYi91}.

It has previously been suggested (Thomson, Chuang and Lin 1986;
Sieradzki and Cammarata 1994; Cammarata and Sieradzki
1996)\nocite{ThChLi86,SiCa94,CaSi96} that surface stresses may influence
dislocation emission from {\em sharp\/} cracks.  However, the surface
stress effect discussed here is of a different nature.  In the case of
a sharp crack the suggested effect of a tensile surface stress would
be to inhibit the emission of dislocations, whereas the tensile
surface stress here is seen to {\em enhance\/} dislocation emission
from blunt cracks.

Gumbsch (1995)\nocite{Gu95} has studied sharp and blunt cracks in Ni
using the Embedded Atom Method (EAM), a many-body potential very
similar to EMT.  Surprisingly, he finds that the sharp crack with the
$(111)[2\bar{1}\bar{1}]$ orientation studied here is stable, and
propagates by cleavage, although other orientations favor dislocation
emission.  The blunt crack is studied only in the $(100)[010]$ and
$(100)[011]$ orientations.  In the first case emission of the usual
${a \over 6}\langle 211 \rangle$ Shockley partial dislocations is
geometrically impossible, and the crack is brittle at all levels of
blunting.  The $(100)[011]$ crack {\em can} emit partial dislocations,
this is seen but only at five layers of blunting, and at a stress
significantly above the Griffith stress (1.17 $K_G$).  The different
behavior in this work may be due to a different crack tip geometry,
since Gumbsch chose a symmetrical configuration, or it may be due to
different surface stresses when the EAM potential is used.

The result that some of these f.c.c.\ metals (Au, Pt) are
intrinsically brittle, and some (Ag, Pd) are close to being
intrinsically brittle, is surprising in light of the "$\gus$"
criterion for brittle/ductile behavior proposed by Rice
(1992)\nocite{Ri92}.  According to that criterion a crack will be
intrinsically ductile if
\begin{equation}
  \label{eq:rice}
  {\gus \over \gsurf} < {2 \over Y}
\end{equation}
where $\gus$ is the ``unstable stacking energy'' (Rice
1992)\nocite{Ri92}, $\gsurf$ is the surface energy and $Y$ is a
geometrical factor.  In our geometry $2/Y$ is approximately 0.30.
$\gus / \gsurf$ is listed in table \ref{tab:emt} for all the metals
studied.  It is seen that by this criterion all the metals should be
intrinsically ductile, and far from the transition to brittle
behavior.  It has been suggested that the stress fields near the crack
tip changes $\gus$ locally.  However this change would result in a
lower value of $\gus$ and thus in enhanced ductility (Sun, Beltz and
Rice 1993)\nocite{SuBeRi93}.  A similar reduction was found by Cleri,
Yip, Wolf and Phillpot (1997)\nocite{ClYiWoPh97}, who used
atomic-scale simulations to calculate the relevant $\gamma$ describing
the barrier for dislocation emission, and found that it was
approximately ${1 \over 2} \gamma_{us}$ for the Lennard-Jones solid
they studied.  We must therefore conclude that the results presented
here are in clear violation of the $\gus$ criterion.

Such a violation
is consistent with previously obtained results for the behavior of
cracks loaded in mode I (pure opening) (Zhou {\em et al.}
1994; Gumbsch and Beltz 1995)\nocite{ZhCaTh94,GuBe95}.  
Eq.~(\ref{eq:rice}) does not consider the energy cost of creating
extra surface at the crack end during dislocation emission.  When this
so called ledge energy is taken into account a different criterion,
where dislocation emission is more difficult, is obtained (Zhou {\em
  et al.\/}~1994)\nocite{ZhCaTh94}.  The resulting criterion is
independent of the surface energy, and depends only on the unstable
stacking energy.  For a two-dimensional hexagonal lattice they find
that a material is ductile if
\begin{equation}
  \label{eq:zct}
  {\gus \over \mu b} < 0.012,
\end{equation}
but the numeric constant may be different in other crystal structures
and other crack geometries.  We have also evaluated $\gus / (\mu b)$
for these materials, see table \ref{tab:emt}.  It is clearly seen that
no value of $\gus / (\mu b)$ can be found that divides the
intrinsically brittle from the intrinsically ductile materials.

As in the case of the blunt cracks, the surface stresses may be
important.  The two metals that appear brittle in this study, Pt and
Ag, are the two metals with the highest tensile surface stress.  This high
surface stress is also seen experimentally, as the surfaces tend to
reconstruct in a way that relieves the surface stress.  Even the
closed-packed (111) surfaces of Pt and Au reconstruct in a way that
increases the density of surface atoms (see e.g.\ Sandy, Mochrie,
Zehner, Gr\"ubel, Huang and Gibbs (1993) and references
therein)\nocite{SaMoZeGrHuGi93}.

When a dislocation is emitted from a sharp crack in the geometry shown
in Figure~\ref{fig:orientation}a, the lower crack surface is stretched
and work is done against the surface stress (Sieradzki and Cammarata
1994)\nocite{SiCa94}.  This work is not included in equation
(\ref{eq:zct}).  The effect of including the surface stress can be
estimated from the work of Thomson and Carlsson (1994)\nocite{ThCa94},
where they give theoretical arguments for equation (\ref{eq:zct}).
They find that the critical value of the crack extension force ${\cal
  G}_{Ie}$ at emission of a dislocation is
\begin{equation}
  \label{eq:emissionold}
  Y^2 {{\cal G}_{Ie} \over \mu' b}  = {\gus \over \mu' a} + 8 \pi
  {\gsurf \gus \over (\mu' a)^2}
\end{equation}
where $Y$ is a geometric factor, $\mu'$ is an effective shear modulus,
$b$ is the Burgers vector and $a$ is the lattice constant.  The
critical crack extension force for cleavage is ${\cal G}_{Ic} = 2
\gsurf$.  The first term in equation (\ref{eq:emissionold} is the work
done to create the dislocation, if it dominates equation
(\ref{eq:rice}) becomes the relevant ductility criterion.  The second
term is the work done to create the ledge, if it dominates equation
(\ref{eq:zct}) determines the behavior.

The work done against the surface stress can be included by replacing
$\gsurf$ in equation (\ref{eq:emissionold}) by $\gsurf + \alpha f$,
where $\alpha$ is the ratio between how much the crack surface is stretched
($u_{\text{tip}}$) and the width of the created ledge.  The work done
does not depend on how the strain is distributed:
\begin{equation}
  \label{eq:work}
  W_{\text{stress}} = \int_{-\infty}^0 dx \, f \varepsilon(x) =
  \int_{-\infty}^0 dx \, f {\partial u \over \partial x} =
  \int_{0}^{u_{\text{tip}}} du \, f =  u_{\text{tip}} f
\end{equation}

The transition between ductile and brittle behavior happens for ${\cal
  G}_{Ie} = {\cal G}_{Ic}$.  Neglecting the first term in equation
(\ref{eq:emissionold}) we get a modified criterion for ductility:
\begin{equation}
  \label{eq:modifcrit}
  {\gus \over \mu b} \left( 1 + \alpha {f \over \gsurf}\right) < C,  
\end{equation}
where $C$ is a constant.  This criterion has been included in
table~\ref{tab:emt}, setting $\alpha = \cos 70.5^\circ = 1/3$ (the
angle between two \{111\} planes).  We see that it is in
\emph{qualitative} agreement with the simulations: the largest values
are found for the metals with brittle behavior (Au and Pt), somewhat
lower values for the metals showing a weak tendensy towards
brittleness (Ag and Pd), and lowest for Cu and Ni.

Equation \ref{eq:modifcrit} should not be taken as a
\emph{quantitative} criterion, as too many assumptions and
approximations were involved.  For example, the two terms in equation
(\ref{eq:emissionold}) are approximately equal in these simulations,
so the criterion should be an intermediate between equations
(\ref{eq:rice}) and (\ref{eq:modifcrit}).  Some of the work done
against the surface stress will be recovered by a contraction of the
opposite surface, reducing $\alpha$ (in the case of emission in the
crack plane $\alpha = 0$ by symmetry).  On the other hand, the cost of
creating the ledge will be less than the surface energy times the
area, as the atoms in the ledge surface have not lost as many
neighbors as in a plane surface.  This would increase the relative
importance of the surface stress versus the surface energy, i.e.\ 
effectively increase $\alpha$.  Although equation \ref{eq:modifcrit}
is not a quantitative ductility critetion, it still provides a
qualitative explanation of the trends observed in the simulations.

It should finally be noticed that the observed intrinsic brittleness
could be an artifact of the low surface energies obtained in the
Effective Medium Theory's descriptions of metallic bonding, although
this may to some extend be compensated by a similar underestimation of
the unstable stacking energy.  However, the test of the ductility
criteria is ``self-consistent'' in the way that the values for $\gsurf$
and $\gus$ are evaluated using the same potential that was used in the
simulations.  In that way even if the values are not exactly the same as
the experimental values, the criteria are still tested reliably.
Similar considerations apply to the blunt cracks, as both the surface
energies and stresses are underestimated.  However, since the ratio
between them is approximately correct, the magnitude of the effect of
crack blunting should be approximately correct.

\begin{center}
  {\sc \S 5. Conclusions}
\end{center}

We have simulated the emission of infinite straight dislocations from
crack tips in Ni, Cu, Pd, Ag, Pt and Au modeled by the Effective
Medium Theory.  We find that a sharp crack is stable and will
propagate in Pt and Au, but is unstable in Ni, Cu, Pd and Ag.  Thus,
gold and platinum are intrinsically brittle whereas the other metals
are intrinsically ductile.  This behavior is not explained by
two recently proposed criteria for intrinsic crack tip behavior. 
The deviations appear to be caused by stresses in the crack
surfaces.

When the cracks are blunted dislocation emission is seen to occur in
all the investigated materials at stresses approximately 20\% below
the theoretical cleavage stress.  We propose that this is caused by a
tensile surface stress in the end-surface of the crack generating a
local stress field that favors emission.  This is supported by a
strong correlation between the change in critical load when the crack
blunting is increased and the strength of the surface stress relative
to the stress from the crack tip.  This effect is not expected to
depend strongly on the details of the geometry of the crack tip.

A consequence of this enhanced emission is a new possible mechanism
for ductile behavior of otherwise intrinsically brittle materials
(Mesarovic 1997)\nocite{Me97}.  In such materials crack blunting,
caused e.g.\ by collisions between the moving crack tip and
preexisting dislocations, could arrest the crack and cause it to emit
dislocations instead.  Such crack arrest has been shown to be able to
cause a dramatic increase in the fracture toughness of the material
(Beltz \emph{et al.} 1996; Mesarovic 1997)\nocite{BeRiShXi96,Me97}.

Since most metals exhibit tensile surface stresses, this mechanism may
be active in other intrinsically brittle materials, and may be a cause
of increased fracture toughness in a class of materials.  Even if the
intrinsic brittleness that we observe for gold and platinum turn out
to be an artifact of the low surface energies in the Effective Medium
Theory, other metals may be intrinsically brittle but be observed to
be ductile because the cracks are arrested by collisions with
preexisting dislocations.

\begin{center}
  {\sc Acknowledgments}
\end{center}

The authors would like to thank Vijay Shastry, Robb Thomson and Karsten W.
Jacobsen for many useful discussions.  Part of this work was supported by the
Department of Energy under grant no.~DE-FG02-84ER45130.  The Center for
Atomic-scale Materials Physics is sponsored by the Danish National Research
Foundation.

\clearpage


\end{document}